\documentclass[preprint,preprintnumbers,aps,prl]{revtex4}

\makeatother

\usepackage{bm}
\usepackage{graphicx}
\usepackage{amssymb}
\usepackage{amsmath}
 
\newcommand{\BEA}{\begin{eqnarray}}
\newcommand{\BEAN}{\begin{eqnarray*}} 
\newcommand{\EEA}{\end{eqnarray}}
\newcommand{\EEAN}{\end{eqnarray*}}

\begin{document}

\title{The Molecular Oxygen Tetramer: Intermolecular Interactions and Implications
  for the $\epsilon$ Solid Phase}

\author{Massimiliano Bartolomei}\email{maxbart@iff.csic.es}
\author{ Estela Carmona-Novillo, Marta I. Hern\'{a}ndez, 
Jes{\'u}s P{\'e}rez-R{\'\i}os, Jos\'e Campos-Mart\'{\i}nez}
\affiliation{Instituto de F\'{\i}sica Fundamental, 
Consejo Superior de Investigaciones Cient\'{\i}ficas (IFF-CSIC), Serrano 123, 
28006 Madrid, Spain}
\author{Ram{\'o}n Hern\'{a}ndez-Lamoneda}
\affiliation{Centro de Investigaciones Qu\'{\i}micas, Universidad 
Aut\'onoma del  Estado de Morelos, 62210 Cuernavaca, Mor. M\'exico}

\date{\today}

\begin{abstract}
 Recent data have determined that the structure of the 
high pressure $\epsilon$ phase of solid oxygen consists 
of clusters composed of four O$_2$ molecules. 
This finding has opened the question about the nature of the
intermolecular interactions within the molecular oxygen tetramer. 
We use multiconfigurational ab initio calculations
to obtain an adequate characterization of the ground singlet state of
 (O$_2$)$_4$  which is compatible with the  non magnetic
character of the $\epsilon$ phase.  In contrast to previous suggestions implying chemical bonding, 
we show that (O$_2$)$_4$ is a van der Waals like cluster where
exchange interactions preferentially stabilize the singlet
state.  However, as the cluster shrinks, there is an extra
stabilization due to many-body interactions that yields a
significant softening of the repulsive wall.
We show that this short range behavior is
a key issue for the understanding of the structure of $\epsilon$-oxygen.

\end{abstract}

\maketitle

The nature of the bonding in molecular oxygen clusters has been a subject of
debate for nearly a century starting with the suggestion by Lewis
of formation of dimers to explain the 
deviations from  Curie's law observed in liquid oxygen\cite{Lewis:24}. 
A chemically bound dimer was expected considering the open shell
$^3\Sigma^-_g$  character of O$_2$ where two unpaired electrons
occupy degenerate $\pi_g^{\star}$ orbitals.
However, a number of experimental\cite{Long:73,Campargue:00,Aquilanti:99} and
theoretical\cite{Wormer:84,maxpccp:08,O2O2pes:10}
works clarified that in the gas phase (O$_2$)$_2$ has the typical 
features of a van der Waals complex: a well depth of tens of meV 
and retention of the molecular properties within the complex.  
In fact it has been shown\cite{Wormer:84,maxpccp:08} that a
singlet species of $D_{2h}$ symmetry is stabilized due to 
exchange interactions but not in a sufficient extent to 
lead to chemical bonding. On the other hand, formation of a O$_4$ molecule  
 with four equivalent single bonds has been theoretically
predicted\cite{Ramon:07} 
but roughly  4.3 eV
above its (O$_2$)$_2$ van der Waals counterpart. This
O$_4$ molecule is structurally closer to the naturally occurring
sulfur rings $S_8$ and in fact an analogous crown-shaped $O_8$ cluster has
been predicted\cite{politzer:00} but again, as a very high energy isomer.     

Recently, the determination of the structure of the  high
pressure  $\epsilon$ phase  of oxygen\cite{o8nature:06,o8prl:06}
has risen the interest in the study of molecular oxygen
oligomers.  In contrast to previously proposed 
structures based on the dimer\cite{Bini:99} and herringbone
chains\cite{na:02}, two independent x-ray diffraction
experiments\cite{o8nature:06,o8prl:06} definitively concluded that
$\epsilon$-O$_2$  consists of layers of well-defined (O$_2$)$_4$
aggregates. They were found to form prisms with  
the O$_2$ axes perpendicular\cite{o8nature:06} or nearly
perpendicular\cite{o8prl:06} to a rhombic, nearly squared, base. 
There is a hierarchy of distances in this phase, the O$_2$ bond 
length which nearly keeps the gas phase value 
($\approx$ 1.21 \AA) at  all pressures, 
and the intra- and inter-cluster distances 
(2.34 and 2.66  \AA, at 11.4 GPa), which decrease 
monotonically with pressure up to the boundary with the 
metallic $\zeta$-phase\cite{o8prl:06}.
Other key properties of this phase, suggesting  increasing intermolecular
interactions, are a dark-red color, a strong 
infrared absorption and a magnetic
collapse\cite{freiman:04,Bini:99,Akahama:00,PRLO8:05}. 
Further evidence for a new intermolecular bonding
has come from inelastic x-ray scattering\cite{PNAS:08} where, at the 
lower pressure boundary of this phase (10 GPa), a discontinuous
shift of about 1.1 eV in the electronic transitions from 1$s$ to 
1$\pi_g^{\star}$ orbitals was found.

 A few works\cite{o8nature:06,O8angchem:07,o8nature:06bis,PhysRevBEpsilon:07}
 have attempted to rationalize the stability and the bonding of the (O$_2$)$_4$ 
 species in the framework of the density functional theory (DFT), but with
 unclear results. Thus, authors of 
 Ref.\cite{O8angchem:07} found that the D$_{4h}$ cuboid structure 
 corresponds to a local energy minimum that they recognized as unstable when
 higher levels of theory were applied.  Furthermore, DFT calculations
 in Ref.\cite{o8nature:06bis} failed to show that the experimental (O$_2$)$_4$
 geometry is the most stable one compared with other chain
 structures\cite{na:02}, showing the need for  additional  studies.  
Then, it is apparent that despite recent 
  progress\cite{becke:07,johnson:10,savin:09} made 
  in DFT methodologies for treating dispersion
  forces, the multiconfigurational character of molecular 
  oxygen clusters still remains a serious challenge for such techniques.

 We report here high level supermolecular ab initio
 calculations of the (O$_2$)$_4$ cluster. Our goal is  a reliable 
 characterization of the {\em singlet} ground state which is consistent
 with the magnetic collapse\cite{PRLO8:05} and 
 spectroscopy\cite{freiman:04,Bini:99,Akahama:00} of the 
 $\epsilon$ phase. To this end, we use a multiconfigurational
 ansatz which is unavoidable for spin multiplicities of (O$_2$)$_4$ lower 
than the maximum one (nonet). We proceed in analogy to our previous work on 
 the dimer\cite{maxpccp:08,O2O2pes:10} and treat the highest spin
 complex by means of   
 a restricted coupled cluster theory with singles,
 doubles and perturbative triple excitations [RCCSD(T)]. 
 In addition, the singlet-nonet splitting can be well described 
 at the multiconfigurational complete active space 
 second order perturbation (CASPT2) theory.
 Finally, the (O$_2$)$_4$ singlet energy is obtained
 by adding, to the RCCSD(T) nonet potential, the singlet-nonet CASPT2
 splitting. The aug-cc-pVQZ\cite{Dunning} basis set has been
used at all levels of theory. For the CASPT2 calculations the active space is
defined by distributing 8 electrons in 8 molecular orbitals correlating 
asymptotically with the O$_2$ $\pi_g^{\star}$ shell. 
As customary, these orbitals have been previously optimized 
with the Complete Active Space Self Consistent Field (CASSCF) method.  
The counterpoise  method \cite{Lenthe:87} was
applied to correct interaction energies for the basis set superposition error. 
As for the  (O$_2$)$_4$ geometry, we use a cuboid structure with $D_{4h}$
symmetry. The centers of mass of O$_2$ form a square whose side
is changed in the range $[1.5-25]$  \AA\ while the intramolecular distance
is kept fixed  to 1.2065\, \AA. 
Calculations have been performed with the {\sc Molpro}2006.1
package\cite{MOLPRO_brief}.  In addition and in order to study the role of many
body interactions, we compare the supermolecular calculations 
just described with estimations based on the summation of 
pure pair interactions. 
In the pairwise approach we have obtained expressions for the
(O$_2$)$_4$  energy which are compatible with a well-defined 
total spin of the complex, resulting in specific combinations of 
the (O$_2$)$_2$ singlet, triplet  and quintet potentials. 
For a faithful comparison, the pair potentials were obtained at the same level
of theory than those of (O$_2$)$_4$. 
It must be noted that there are three singlet states\cite{PhysRevBO8:07} 
asymptotically correlating with four O$_2$($^3\Sigma^-_g$), and that 
here we are reporting the calculations of the {\em ground} singlet state.

\begin{figure}[t]
\includegraphics[width=8cm,angle=0.]{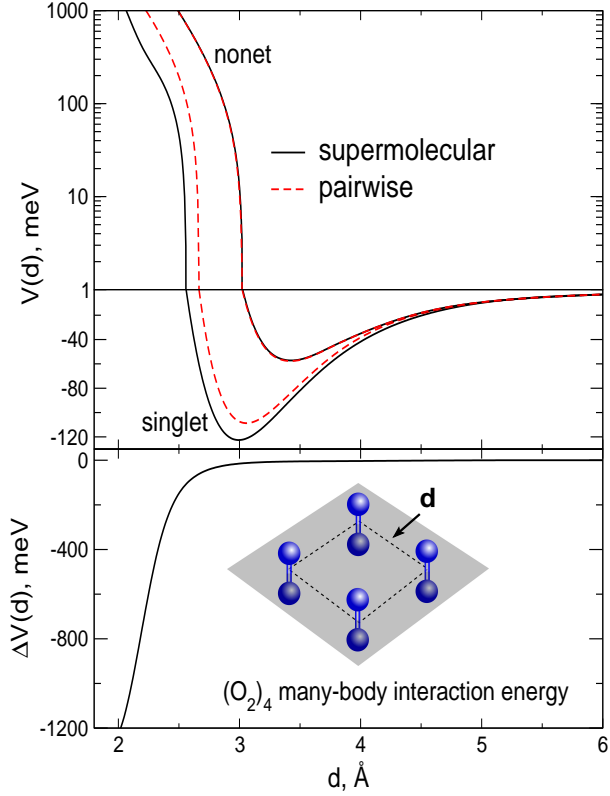}
\vspace*{.1cm}
\caption[]{
Upper panel: Interaction energies (in meV) for  the
 nonet and ground singlet states of (O$_2$)$_4$ as 
 functions of the square side $d$ (in \AA\ ). Supermolecular approach is 
represented in solid lines while the pairwise approach in dashed lines. 
 Lower panel: many-body interaction energy for the singlet state, $\Delta
 V(d)$, obtained as
 the difference between supermolecular and pairwise energies. The large values
of $\Delta V$ at small $d$'s might be a key feature to explain the clustering
of O$_2$ molecules in the $\epsilon$ phase.}
\label{fig1}
\end{figure}

\begin{table}
\vspace*{-.5cm}
\caption{ 
Binding energy  $D_e$ (in meV) and equilibrium distance $R_e$  (in \AA)
of the (O$_2$)$_4$  potentials reported in Fig. 1}
\label{table1}
\renewcommand{\arraystretch}{0.8}
\tabcolsep0.1cm
\begin{center}
\begin{tabular}{lcccccc}
       &    &  \multicolumn{2}{c}{nonet}  & & \multicolumn{2}{c}{singlet} \\
\hline
       &    &  $D_e$   &  $R_e$  & & $D_e$  &  $R_e$ \\
 supermolecular & & 56.6 & 3.37 & &  122.6 &  2.99 \\
 pairwise       & & 57.1 &  3.37 & & 108.4 &  3.08 \\
\hline
\end{tabular}
\end{center}
\vspace*{-.4cm}
\end{table}

The interaction energies of (O$_2$)$_4$  in the singlet and nonet states as 
functions of the square side $d$ are reported in the upper panel of Fig. 1
together with the pairwise estimations of the corresponding interactions. 
Equilibrium parameters of these potentials are  given in
Table I. As can be seen from the parameters of the singlet and
nonet potential  wells, (O$_2$)$_4$ is a van der  Waals like
complex in the gas phase, mainly stabilized by dispersion interactions. 
Exchange interaction, however, 
plays  a role making the potential well of the singlet state deeper and
shifted at shorter  intermolecular distances than that of the
nonet state.  
As in the dimer\cite{maxpccp:08,O2O2pes:10}, the exchange interaction
favors the states of lowest spin multiplicity.  

More insight is gained when comparing the supermolecular calculations with the
pairwise estimations. For the nonet state, the pairwise
approximation reproduces very well the supermolecular energies
indicating that many-body effects are not particularly relevant.
 However, for the singlet state, an analogous
agreement is only achieved for the larger $d$  sizes of the cluster.
Around the minimum of the singlet well, the supermolecular
energies are already lower than the pairwise ones (see
Table I) but it is for shorter distances where
a remarkable softening of the repulsive wall is found. This is due to a
many-body effect, as shown in the lower panel
of Fig. 1 where it can be noticed that the many-body interaction energy for the singlet state
 increases dramatically as $d$ decreases, being 
about 1.2 eV at $d$= 2 \AA. 
The origin of this effect must be in the exchange interactions\cite{Kaplan:11} since
polarization contributions to many-body interaction energies, if important,
should be shown also for the nonet state, that is not the case.
 
In a molecular crystal the first response to pressure is
the ``squeezing out of van der Waals space'', in other words, 
the penetration to the repulsive region of the intermolecular 
potentials\cite{hoffmann:07}. Therefore the peculiar behavior reported above
 can be relevant indeed for understanding the structure of the high
pressure $\epsilon$ phase\cite{o8nature:06,o8prl:06}. For
this reason, in the following we will focus on the features 
of (O$_2$)$_4$ at short cluster $d$ sizes.

\begin{figure}[t]
\includegraphics[width=7.0cm,angle=0.]{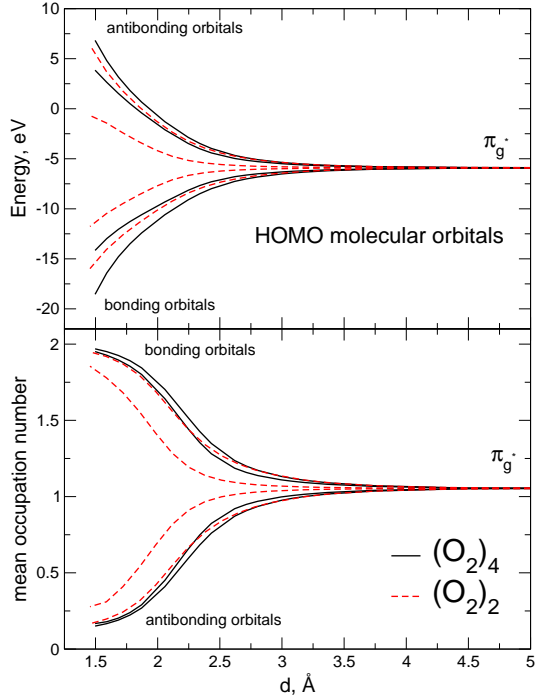}
\caption[]{
Upper panel: Energies of the HOMOs originated from 
the O$_2$ $\pi_g^{\star}$ orbitals vs. intermolecular distance
$d$, for the  ground singlet states of (O$_2$)$_4$  and (O$_2$)$_2$. 
Note that for the tetramer each
line corresponds to a couple of almost degenerate HOMOs. 
Lower panel: As in the upper panel but for occupation numbers.} 
\label{fig2}
\end{figure}

Some clues for the softening of the repulsive wall of the ground singlet
state are given in Fig. 2, where we report CASSCF calculations of the
molecular orbitals arising from the interaction of the eight half-occupied
$\pi_g^{\star}$ orbitals of O$_2$. An analogous calculation for the singlet
(O$_2$)$_2$ is shown for the sake of comparison. The properties of these
optimized orbitals barely change from the asymptote up to 
the equilibrium distance ($d \approx 3$
\AA), but for shorter $d$'s the interaction does give rise to bonding and
antibonding orbitals. Interestingly, the four bonding orbitals in (O$_2$)$_4$
are more stable than the bonding orbitals of (O$_2$)$_2$ and the associated
occupation numbers increase faster as $d$ decreases, as a result of
many-body exchange effects.  However, a double
occupancy only occurs for very short distances  ($d<$ 1.8 \AA). Moreover,
the (O$_2$)$_4$ orbital stabilization is much smaller than the electron-electron Coulomb
repulsion contribution to the total interaction. Thus, the result is not a
minimum but a softening of the repulsive region of the potential, as obtained
in Fig. 1. This multiconfigurational analysis differs from those of
Refs. \cite{O8angchem:07,PNAS:08} where, based on a simpler
monoconfigurational picture, it was suggested that all the bonding orbitals
were doubly occupied leading to large binding energies with respect to the
isolated O$_2$ molecules. 
The present analysis also gives a qualitative insight
into the observation of a $\approx$ 1 eV shift in the $\pi_g^{\star}
\leftarrow$ 1 $s$ transitions at the boundary of the
$\epsilon$ phase\cite{PNAS:08}, since the splitting between the
(O$_2$)$_4$ antibonding energies and the isolated  $\pi_g^{\star}$ orbitals is
of the same order of magnitude ($\approx$ 1.5 eV) in the relevant range ($d
\approx$ 2.4 \AA).

\begin{figure}[t]
\includegraphics[width=8.cm,angle=0.]{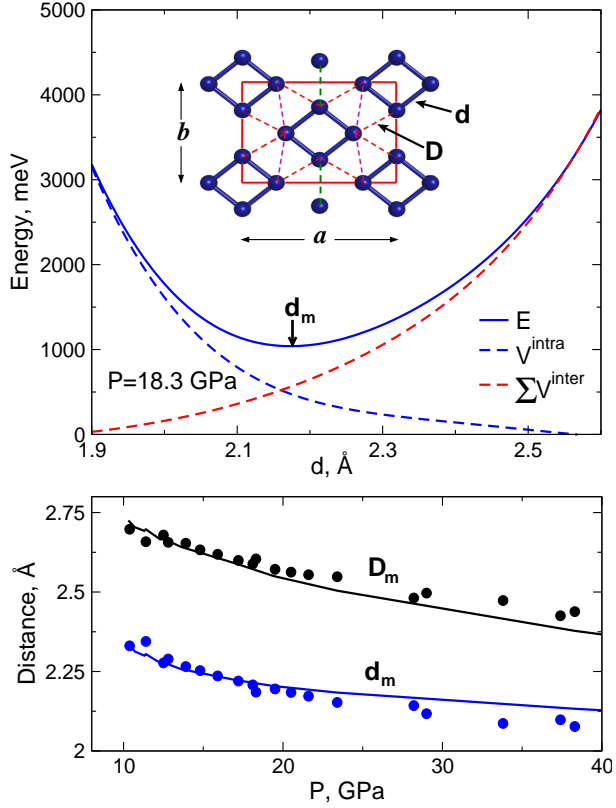}
\caption[]{Upper panel: Energy of a layer of $\epsilon$-O$_2$ (Eq. 1) as
  a function of the cluster size, $d$, for a pressure of 18.3 GPa (solid
  line). In the inset, the model unit cell is shown  where the
  O$_2$ axes are perpendicular to the $a$-$b$ plane and the inter-cluster
  distances, $r_{ij}$, are displayed by dashed lines, being $D$ the shortest
  one among them. 
  A minimum of the energy is obtained for the intra-cluster distance $d_{m}$. 
  Lower panel: Pressure dependence of $d_{m}$ and $D_{m}$ (lines)  compared with data
  of Ref.\cite{o8prl:06}(circles). See text for details.
}
\label{fig3}
\vspace*{-.2cm}
\end{figure}

 We have also checked that the energy of the ground singlet state 
is significantly lower than those of the other singlet states and 
different spin multiplicities as well, particularly in the 
repulsive region of the interaction. 
Albeit in a very intuitive manner, the emerging picture can give some 
hints on the formation and structure of the $\epsilon$ phase and 
its non-magnetic character. 
To illustrate this point and also in order to study whether present 
results are adequate for a description of the $\epsilon$ phase,   
we consider a very simple model for the energy of a layer of (O$_2$)$_4$
clusters. A basic unit cell is shown in Fig.\ref{fig3} where the
tetramers form rhombuses of length $d$ and angle $\alpha$. It is assumed that,
at a given pressure, the centers of mass of the clusters are fixed and
determined by the lattice parameters $a$ and $b$. Values of these parameters
as functions of pressure were taken from Ref.\cite{o8prl:06}. In addition, the
angle $\alpha$ is fixed to  81.4$^{o}$ as derived from data reported at
11.4 GPa\cite{o8prl:06}. We assume that all clusters increase/decrease their
size $d$ at a time and study the subsequent modification of the cell
energy. This energy is given as a sum of intra- and inter-cluster
contributions   

\vspace{-.3cm}

\begin{equation}
 E(d)  =  V^{intra}(d)+ \frac{1}{2}\sum_{i,j}V^{inter}_{ij}(r_{ij}), 
\label{epsilon}
\end{equation}
\vspace{-.3cm}

\noindent
where $V^{intra}$ is the already reported supermolecular (O$_2$)$_4$ singlet
potential and $V^{inter}_{ij}$ is a pair potential between molecules $i$ and $j$
belonging to different clusters. The corresponding intermolecular distance
$r_{ij}$ is determined by $d$, $a$, $b$ and $\alpha$. A spin-averaged 
(O$_2$)$_2$ potential was used for $V^{inter}_{ij}$ because the nonmagnetic
character\cite{PRLO8:05} of the $\epsilon$ phase suggests that dependence on
spin must be washed out. In Fig.\ref{fig3} (upper panel) it is shown that,
within the cell, the optimum size of the cluster is considerably reduced with
respect to the gas phase equilibrium distance. For 18.3 GPa, a minimum in the
total energy is obtained at about $d_{m}$= 2.17 \AA, in agreement with the
observed intra-cluster distance of 2.185 \AA\cite{o8prl:06} as well as with the
value obtained at 17.6 GPa in other independent
experiment\cite{o8nature:06}. 
In the lower panel of Fig. 3 we show that the pressure dependence of
the optimized intra- and inter-cluster distances $d_m$ and $D_m$ agrees fairly
well with the observations\cite{o8prl:06} ($D_m$ is the shortest inter-cluster
distance, obtained as a function of $d_m$, $a$, $b$ and $\alpha$). 

We would like to stress that, despite the simplicity of the model, the key
element is the behavior of the repulsive wall of the ground singlet state, 
adequately calculated at a multiconfigurational level of theory.  
Indeed, substitution of the the ground singlet
  energy $V^{intra}(d)$ with that corresponding to a different
spin multiplicity (or an excited singlet state) would lead to an optimum
intracluster size far less compatible with the measurements (e.g., $d_m
\approx$ 2.35-2.40 \AA\ at 18.3 GPa). As discussed in Ref.\cite{o8nature:06}, 
the fact that both intra- and inter-cluster distances compress at nearly the
same  rate is a clear indication of a rather weak interaction between the O$_2$
molecules. Present finding of a van der Waals cluster with a very 
incipient chemical bond for short sizes is consistent with the
observations. 
More refined models and extended calculations should be developed to account
for the intriguing spectroscopy of
$\epsilon$-oxygen\cite{freiman:04,Bini:99,Akahama:00}. 
Since at the cluster sizes relevant to the $\epsilon$ phase the energy of the 
(O$_2$)$_4$ unit is not a minimum,   
 including both intra and inter-cluster degrees of freedom in such models is
 unavoidable, especially for the lower frequency vibrational modes.

We thank Yuichi Akahama for sending us the structure data of
Ref.\cite{o8prl:06} and funding 
 by Ministerio de Ciencia e Innovaci{\'o}n (Spain, 
  FIS2010-22064-C02-02). RHL was supported by a
 sabbatical grant by the Ministerio de Educaci{\'o}n (Spain, SAB2009-0010)
 and by Conacyt (Mexico, Ref. 126608).  
J.P.-R. is a JAE CSIC predoctoral fellow.

\end{document}